\documentclass[useAMS,usenatbib,usegraphicx]{mn2e}
\usepackage{amsmath,amsfonts,amssymb}
\usepackage[colour]{optional}

\newcommand{\1}{^{-1}}
\newcommand{\hm}{\mbox{$h^{-1}$}}
\newcommand{\vir}{_{\textrm{vir}}}
\newcommand{\bj}{\mbox{$b_{\rm J}$}}
\def\Msun{\hbox{$\rm\, M_{\odot}$}}
\newcommand{\canonical}{canonical}
\newcommand{\disruption}{disruption}

\title[Disruption of dwarfs]{The effect of dwarf galaxies disruption in semi-analytic models}

\author[Henriques et al.]
{Bruno M. Henriques$^{1}$\thanks{E-mail: b.m.henriques@sussex.ac.uk},
 Serena Bertone$^{1}$
 and Peter A. Thomas$^{1}$ \\
 {}$^{1}$Astronomy Centre, University of Sussex, Falmer, Brighton BN1 9QH,
 United Kingdom}

\begin{document}

\date{Accepted by MNRAS}

\pagerange{\pageref{firstpage}--\pageref{lastpage}} \pubyear{2007}

\maketitle

\label{firstpage}

\begin{abstract}

We present results for a galaxy formation model that includes a
simple treatment for the disruption of dwarf galaxies by gravitational
forces and galaxy encounters within galaxy clusters. This is
implemented \emph{a posteriori} in a semi--analytic model by
considering the stability of cluster dark matter sub-haloes 
at $z=0$. We assume that a galaxy whose dark matter 
substructure has been disrupted will itself disperse, while its stars
become part of the population of intracluster stars responsible for
the observed intracluster light.  Despite the simplicity of this
assumption, our results show a substantial improvement over previous
models and indicate that the inclusion of galaxy disruption is indeed
a necessary ingredient of galaxy formation models. We find that galaxy
disruption suppresses the number density of dwarf galaxies by about
a factor of two. This makes the slope of the faint end of the galaxy
luminosity function shallower, in agreement with observations. In
particular, the abundance of faint, red galaxies is strongly
suppressed. As a result, the luminosity function of red galaxies and
the distinction between the red and the blue galaxy populations in
colour--magnitude relationships are correctly predicted. Finally, we
estimate a fraction of intracluster light comparable to that found in
clusters of galaxies.

\end{abstract}

\begin{keywords}
methods: numerical -- galaxies: formation -- galaxies: evolution -- galaxies: interaction
\end{keywords}

\section{Introduction}
\label{sec:intro}

The existence of a diffuse population of intracluster stars was first
proposed by \citet{zwicky1951} and has since been detected
unambiguously.  Individual intracluster AGB stars \citep{durrell2002},
planetary nebulae (\citealt{Arnaboldi1996}, \citealt{Feldmeier2003},
\citealt{Arnaboldi2004}, \citealt{Feldmeier2004a}), novae and
supernovae (\citealt{Gal-Yam2003}, \citealt{Neill2005}) have been
resolved in nearby clusters and trace an unbound population of red,
old stars orbiting freely in the cluster potential \citep{Krick2006}.

The light associated with intracluster stars, or diffuse intracluster
light (ICL), can contribute between 10 and 40 per cent of
the optical emission of rich galaxy groups and clusters
(\citealt{Bernstein1995}; \citealt{Gonzalez2000};
\citealt{Feldmeier2002}; \citealt{Feldmeier2004b};
\citealt{Gonzalez2005}; \citealt{Zibetti2005}). In the core of massive
clusters like Coma, ICL emission not associated with the central giant
elliptical galaxy can account for up to 50 per cent of the total
optical light \citep{Bernstein1995}.

Rather than having formed in the intracluster medium, intracluster
stars are believed to be the remnants of disrupted galaxies.
Gas-dynamical simulations generally agree that the bulk of the ICL is
emitted by stars that have been continually stripped from member
galaxies throughout the lifetime of a cluster, or have been ejected
into intergalactic space by merging galaxy groups
(\citealt{Moore1996}; \citealt{Napolitano2003}; \citealt{Murante2004};
\citealt{Willman2004}; \citealt{Sommer-Larsen2005};
\citealt{Monaco2006}; \citealt{Murante2007}).  Several mechanisms can
contribute to disrupting galaxies, the most efficient ones being tidal
stripping by the gravitational forces of the cluster halo
\citep{Merritt1984} and high-speed galaxy encounters
\citep{Richstone1976}. Smoothed particle hydrodynamic 
simulations by \citet{Murante2007} show that 60 to 90 per cent of the
population of intracluster stars has been generated at $z<1$ by
relaxation processes during galaxy mergers, with a minor fraction
arising from tidal stripping. In groups, low speed galaxy encounters
and dynamical heating by galaxy scattering with halo substructures may
enhance the removal of stars by tidal stripping once a group merges
with a larger cluster (\citealt{Gnedin2003}; \citealt{Mihos2004};
\citealt{Rudick2006}).

Low surface brightness features have been identified in the Coma and
Centaurus clusters (\citealt{Gregg1998}; \citealt{Trentham1998};
\citealt{Feldmeier2002}), indicating the presence of dynamically-young
tidal structures produced by the disruption of infalling galaxies.
These features carry information about the orbit of the parent galaxy
and can be used to trace the cluster accretion history and its past
dynamical interactions \citep{calcaneo2000}.
Another indication comes from the number density of satellites in groups. 
\citet{Faltenbacher2005} show that the projected number density profile of 
dwarf galaxies in NGC 5044 can only be explained by assuming that a significant 
amount of mass in satellite galaxies is tidally disrupted.

Further compelling evidence for galaxy disruption comes from the population of ultra compact dwarf galaxies in clusters (UCDs, \citealt{Hilker1999};
\citealt{Drinkwater2000}; \citealt{Hilker2007}), which is believed to
form when the low surface density disk component of a dwarf galaxy is
tidally-stripped away. Indeed, UCDs might share a common origin with
intracluster stars (\citealt{Zibetti2004}).

High-resolution numerical simulations suggest that galaxies may
disperse a non-negligible fraction of their stellar component into
the intracluster medium during mergers (\citealt{Monaco2006};
\citealt{Murante2007}; \citealt{Conroy2007}). \citet{Monaco2006} find
that the small degree of evolution seen at $z<1$ for the high-mass end
of the stellar mass function can be accounted for if merging galaxies
eject at least 20 per cent of their stars. The joint analysis of
simulation-based models and observations of ICL by \citet{Conroy2007}
favours models in which as much as 80 per cent of stars are stripped
into the intracluster medium from satellite galaxies whose dark matter
substructure has been disrupted.

If, as simulations suggest, intracluster stars have been stripped from
galaxies with all ranges of masses (\citealt{Bullock2001}; \citealt{Taylor2001}; 
 \citealt{Benson2002}; \citealt{Monaco2006}; \citealt{Murante2007}),
 it is likely that the faint end of the
luminosity and stellar mass functions are affected as strongly as the
bright end.  However, while several works have investigated the effect
of galaxy disruption on the evolution of massive galaxies, there are
only a couple of suggestions in the literature that pinpoint the
effect of disruption on the abundance of dwarfs
(\citealt{Zibetti2004}, \citealt{Zibetti2005}).

In this work, we use simulated galaxy catalogues, produced by
semi-analytic models of galaxy formation\footnote{The catalogues are
publicly available at:\\ http://www.mpa-garching.mpg.de/millennium/
and \\ http://www.virgo.sussex.ac.uk/Millennium/millennium.html}
(\citealt{deLucia2007}, hereafter DLB07; \citealt{Bertone2007},
hereafter BDT07) to investigate how
galaxy disruption affects the abundance of dwarf galaxies by
considering its effects on the faint end of the stellar mass and
luminosity functions.  In particular, we focus on the abundance of
dwarf red galaxies, which is considerably overpredicted by current
galaxy formation models \citep{Croton2006}.

This paper is organised as follows. Section \ref{sec:model} briefly
describes the galaxy catalogues and our method to identify disrupted
dwarf galaxies. In Section \ref{sec:results} we present our results
and in Section \ref{sec:discuss} we draw our conclusions.

\section{The model}
\label{sec:model}

In this Section we briefly describe the Millennium galaxy catalogue we
use for this work and our method for identifying disrupted galaxies in
groups and clusters.

The Millennium Simulation traces the evolution of dark matter haloes
in a cubic box of 500\hm Mpc on a side \citep{Springel2005}. It
assumes a $\Lambda$CDM cosmology with parameters $\Omega_{\rm
m}=0.25$, $\Omega_{\rm b}=0.045$, $h=0.73$, $\Omega_\Lambda=0.75$,
$n=1$, and $\sigma_8=0.9$, where the Hubble parameter is $H_0 = 100$
\hm km s$\1$ Mpc$\1$. The simulation follows $2160^3$ dark matter
particles of mass $8.6\times 10^8$ \hm M$_{\sun}$. Since dark matter
haloes are required to contain at least 20 particles, the minimum halo
mass is $1.7\times 10^{10}$ \hm M$_{\sun}$, with a corresponding
baryonic mass of about $10^{9.5}$ \hm M$_{\sun}$.

Currently, three different galaxy catalogues based on the Millennium
simulation have been publicly released (DLB07; \citealt{Bower2006};
BDT07).  The catalogue of \citet{Bower2006} is not suitable for this
study, because it does not distinguish between cluster galaxies
associated with dark matter substructures and those whose substructure
has completely merged with the parent halo.  The catalogues of DLB07
and BDT07 do make this distinction and are therefore best suited to
our purposes.  These last two models differ only in the adopted scheme
for galactic feedback.  The model of DLB07 uses the empirical feedback
scheme of \citet{Croton2006}, while BDT07 implement a dynamical
feedback scheme that directly solves the equation of motion for
the evolution of galactic winds \citep{Bertone2005}.  Both
models over-predict the abundance of dwarf galaxies, and in particular
that of dwarf red galaxies, by a factor of two (BDT07) to five
(DLB07).  The model of BDT07, thanks to the more efficient feedback,
produces stellar and gas metallicities in better agreement with
observations and correctly predicts the slope of the red sequence in
colour-magnitude diagrams.  The model of DLB07 predicts a rather flat
distribution of red galaxies, which might be due to the high
metallicities predicted for dwarf galaxies.  However, globally the
model of DLB07 predicts galaxy colours that reproduce the bimodality
seen in observations (\citealt{Baldry2004}; \citealt{Baldry2006})
somewhat better. For this reason, in this work we will present results
for the model of DLB07. However, we note that all results described in the following Sections are qualitatively unchanged for the model of BDT07.

The Munich semi-analytic models of galaxy formation (DLB07 and BDT07)
are an ideal instrument for investigating the origin of interstellar
stars.  The substructures of dark matter haloes can be identified by
the algorithm {\small SUBFIND} \citep{Springel2001}.  As a
consequence, galaxies in clusters can be of three different types:
central galaxies, satellites of ``type 1'' and satellites of ``type
2''.  Satellites of type 1 are associated with dark matter
substructures, which usually are recently-merged haloes.  Satellites
of type 2 are instead galaxies whose dark matter halo has completely
merged with a bigger halo and are not associated with a substructure.
In the models, only type 2 galaxies merge with the halo central
galaxy, after a dynamical timescale.  In this work, we include galaxy
disruption in the DLB07 model \emph{a posteriori} and investigate its
effects at $z=0$.  To do this, we assume that type 2 galaxies have
been completely stripped of their stars and that these stars represent
the population of intracluster stars that produces the ICL.  Type 2
galaxies therefore do not contribute to galaxy counts.  These
assumptions are motivated by the fact that type 2 galaxies in the
simulation usually fall into a larger structure at $z>1$ and are not associated with a dark
matter substructure.  This implies that they have had enough time to
experience a number of low and high speed encounters with other
galaxies during their lifetime, and that most of their stars might
have been stripped away when their associated dark matter substructure
was disrupted.

The definition of galaxy type that we adopt is dependent on the
resolution of the dark matter simulation. An increase in resolution
would in principle mean that type 1 haloes survive longer before being
disrupted.  In addition, disruption is a gradual process and one might
expect most type 1 galaxies to be partially stripped, while some type
2 galaxies retain a compact core. We considered implementing partial
disruption, but decided against it mainly because it would introduce
extra parameters into the model without significantly improving the
quality of the results.  Our model should be regarded as a qualitative
indicator of the importance of disruption rather than a detailed
physical model of these processes.  

Although this simple model allows us to investigate the effect of
galaxy disruption on the abundance of dwarfs, it does not implement
the disruption self-consistently within the semi-analytic model.
Since we apply our assumption \emph{a posteriori}, we do not modify the
evolution of type 2 galaxies at $z>0$. Instead, we consider as
disrupted only type 2 galaxies that have have not yet merged at $z=0$.
Our assumption of complete galaxy disruption, if consistently
implemented, would eliminate merging from the model, because in the
Munich semi-analytic model only type 2 galaxies can merge with central
galaxies. A self-consistent treatment of disruption could for example
include partial disruption of the merging galaxies, as assumed by
\citet{Monaco2006}, or follow the forces acting on satellites, to
define a disruption radius, outside which the galaxy mass will be lost 
(\citealt{Bullock2001}; \citealt{Taylor2001}; \citealt{Benson2002}).
We leave this study for future work.

Throughout the rest of this paper, we refer to the original
model of DLB07 as the ``canonical'' model, and the equivalent model
including destruction of type 2 galaxies as the ``disruption'' model.

\section{Results}
\label{sec:results}

In this Section we present our results for the colour-magnitude
diagram (Subsection \ref{sec:cm}) and for the galaxy luminosity
function in \bj-band (Subsection \ref{sec:lf}). Finally, Subsection
\ref{sec:massfrac} shows our predictions for the fraction of
interstellar light produced by disrupted galaxies.

\subsection{The colour-magnitude diagram}
\label{sec:cm}

Colour-magnitude diagrams indicate the existence of two distinct
populations of galaxies, namely a population of blue, star-forming
galaxies and a population of red galaxies with no significant star
formation activity.

\begin{figure*}
\centering \includegraphics[width=16.8cm]{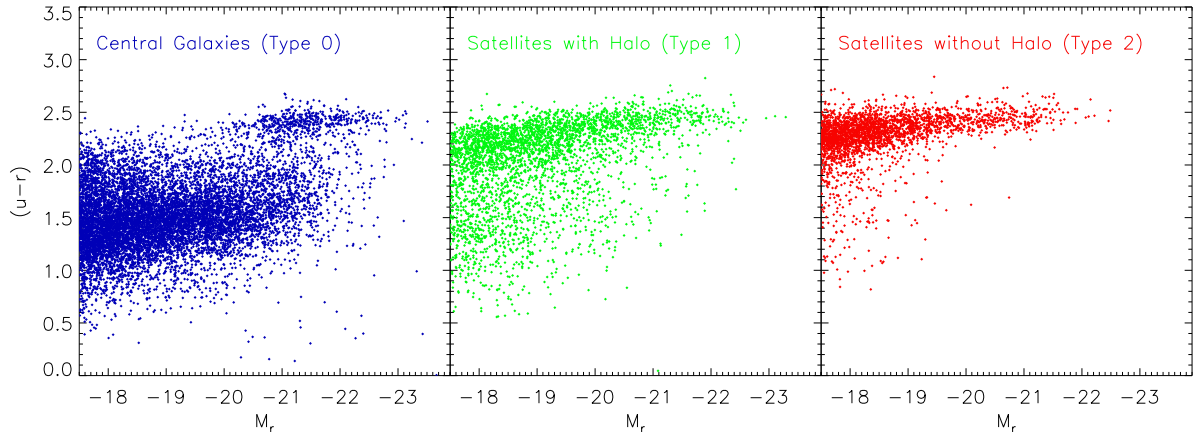}
\caption{The $u-r$ colour distribution of galaxies as a function of
$M_{\rm r}$ for the canonical model. Galaxies have been colour-coded
by type: central galaxies (blue, left panel), type 1 galaxies (green,
central panel) and type 2 galaxies (red, right panel).}
\label{fig:types}
\end{figure*}

Fig.~\ref{fig:types} shows the $u-r$ colour distribution of galaxies
of different types in the \canonical\ model, as a function of $M_{\rm
r}$.  The faint reddest galaxies (in red, top panel) are predominantly
type 2 galaxies. Their colour and the amount of light produced by
their stars agree well with the observed properties of the ICL.  Type
1 galaxies have somewhat bluer colours than type 2 galaxies, because
they might not have yet exhausted their reservoir of gas and may still
be forming stars. However, as can be seen in the bottom panel of
Fig. \ref{fig:types}, a large fraction of type 1 galaxies accumulates
along the red sequence, once their reservoir of gas is extinguished.
Most central galaxies are blue (in blue, bottom panel), because they
can continually accrete gas and form stars. The exceptions are very
massive galaxies, in which star formation is suppressed by AGN
feedback (\citealt{Croton2006}; \citealt{Bower2006}) and a significant
fraction of dwarf galaxies. These last ones mostly form in haloes with
less than about 100 dark matter particles, for which the resolution of
the Millennium simulation is insufficient to properly account for gas
accretion and merging.

\begin{figure}
\centering
\includegraphics[width=8.4cm]{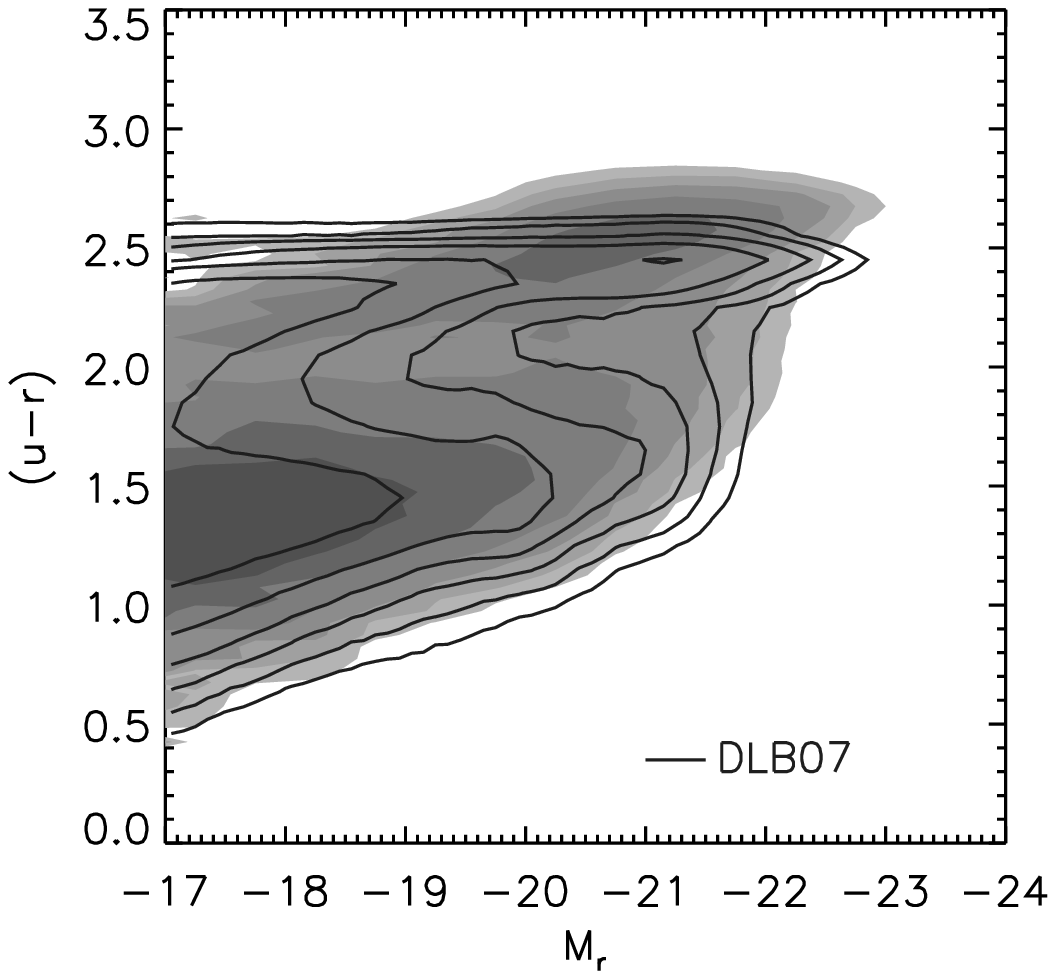}
\includegraphics[width=8.4cm]{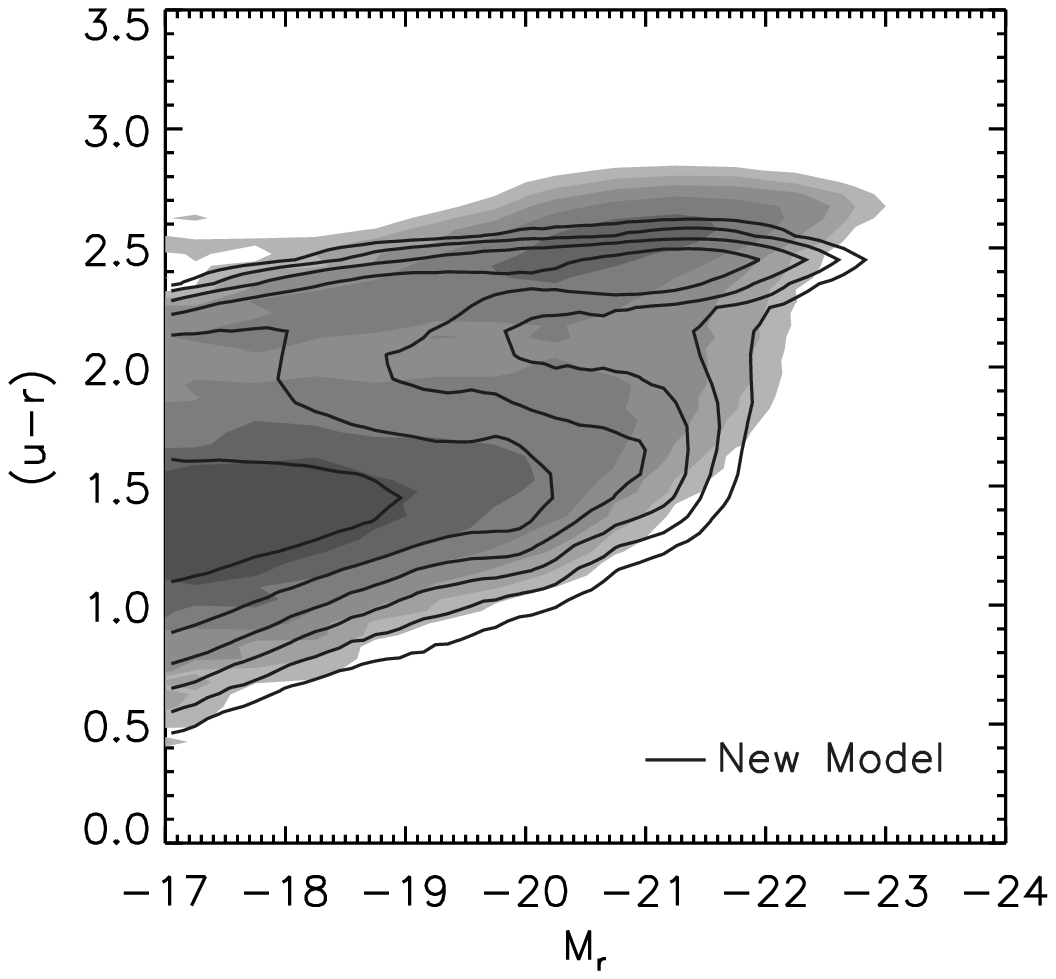}
\caption{Comparison of the colour-magnitude relation for galaxies in
the \canonical\ model (upper panel) and in the \disruption\ model
(lower panel). The shaded contours are for the observed distribution
of \citet{Baldry2004}. The contour scale is logarithmic, with 4
divisions per dex.}
\label{fig:cm}
\end{figure}

\begin{figure}
\centering
\includegraphics[width=8.4cm]{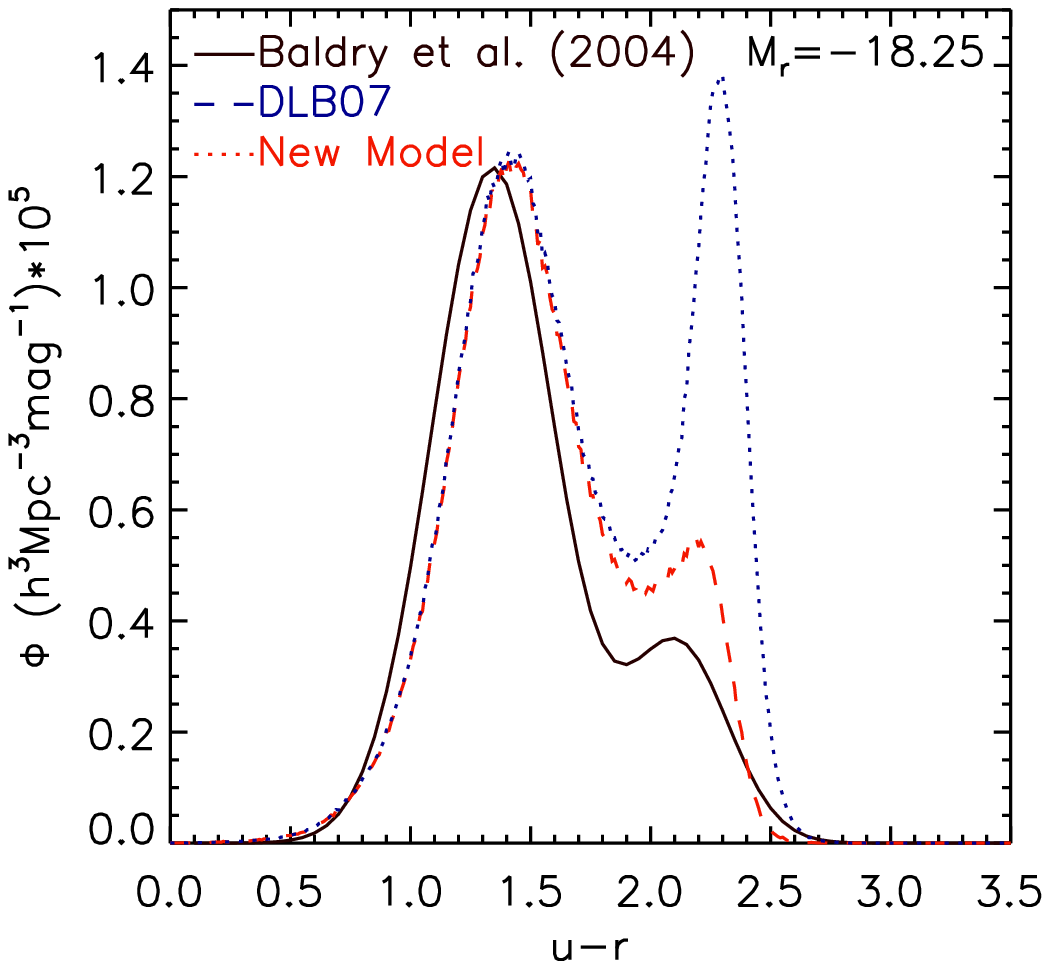}
\caption{The u-r colour distribution for an absolute magnitude bin of
width 0.5 centred at $M_{\rm r}$=-18.25. The predictions for the
disruption and canonical models (red dashed and blue dotted lines
respectively) are compared with observational data from
\citet{Baldry2004} (black solid line).}
\label{fig:chist}
\end{figure}

In Fig.~\ref{fig:cm} we directly compare the colour-magnitude
distribution for the \canonical\ model (top panel) and for the
\disruption\ model (bottom panel) with the observed distribution in
$u-r$ of \citet{Baldry2006}.  The blue population is modelled with a
sufficient degree of accuracy. On the other hand, the red population
considerably differs from the observed one.  Firstly, the colour of
bright, red galaxies is underestimated by about 0.2 magnitudes. This
is likely due to the underestimate of the metallicity of massive
galaxies in the DLB07 model, as shown by BDT07. In fact, both the
BDT07 model and that of \citet{Bower2006} do predict higher
metallicities and redder colours for the brightest galaxies. This may
also explain the flat distribution for the population of red galaxies
in the top panel of Fig. \ref{fig:cm}, in contrast with the sloping
behaviour seen in observations (\citealt{Baldry2004};
\citealt{Baldry2006}) and predicted by other models
(\citealt{Bower2006}; BDT07).

A more serious problem, however, is that the \canonical\ model vastly
over-predicts the abundance of faint red galaxies.  The \disruption\
model, whilst obviously not completely solving the problem, does
produce a distribution of red galaxies with a distinct red peak at
$M_{\rm r} \approx -21$ and a decreasing rather than flat distribution
towards fainter magnitudes. This can be seen in
Fig.~\ref{fig:chist}, where the colour distribution of galaxies with
absolute magnitudes in the range $M_{\rm r}=-18.25 \pm 0.5$, for the
canonical and the disruption models, are compared to the observational
data of \citet{Baldry2004}. In this magnitude bin, the disruption
model predicts a considerably smaller number of red galaxies, in
better agreement with the observations. This is because the reddest
dwarf galaxies are type 2 galaxies, which we assume become disrupted
once their dark matter substructure has been lost.  This clearly
indicates that galaxy disruption might play a role in shaping the
faint side of colour-magnitude distributions.

The DLB07 model that we are using here predicts too small a
fraction of red galaxies in low-density environments \citep[see
e.g.~][figure~15]{Baldry2006}.  Our model preferentially disrupts red
galaxies in clusters and so does not help to correct this deficiency:
in fact, the predicted red fractions are reduced slightly from those
shown in \citet{Baldry2006}. On the other hand, the \citet{Bower2006}
model has a higher red fraction that provides a better fit to the
data.

\subsection{Luminosity functions}
\label{sec:lf}

\begin{figure}
\centering \includegraphics[width=8.4cm]{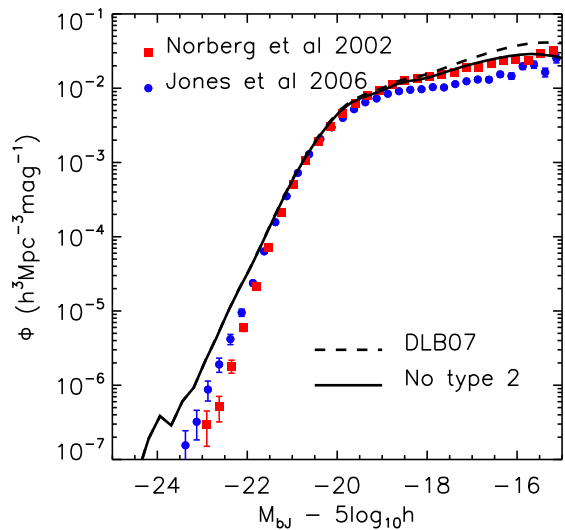}
\caption{The galaxy luminosity function in the \bj\ photometric band
at $z=0$ for the canonical (dashed line) and disruption (solid line)
models. The numerical predictions are compared to the results of the
2dFGRS \citep{Norberg2002} and to those of the 6dFGS
\citep{Jones2006}.}
\label{fig:lf}
\end{figure}

In this Subsection we investigate the effect of disruption on the
abundance of galaxies, and in particular on the luminosity function in
the \bj\ photometric band at $z=0$.  Fig.~\ref{fig:lf} shows the
luminosity functions for the \canonical\ and \disruption\ models
(dashed and solid lines respectively). The data points reproduce the
observational results of the 2 degree Field Galaxy Redshift Survey
(2dFGRS, \citealt{Norberg2002}) and of the 6 degree Field Galaxy
Survey (6dFGS, \citealt{Jones2006}).  The disruption of satellite
galaxies strongly affects the faint end of the luminosity function,
changing the slope from positive to negative, and giving better 
agreement with the observations than does the canonical model.

\begin{figure}
\centering
\includegraphics[width=8.4cm]{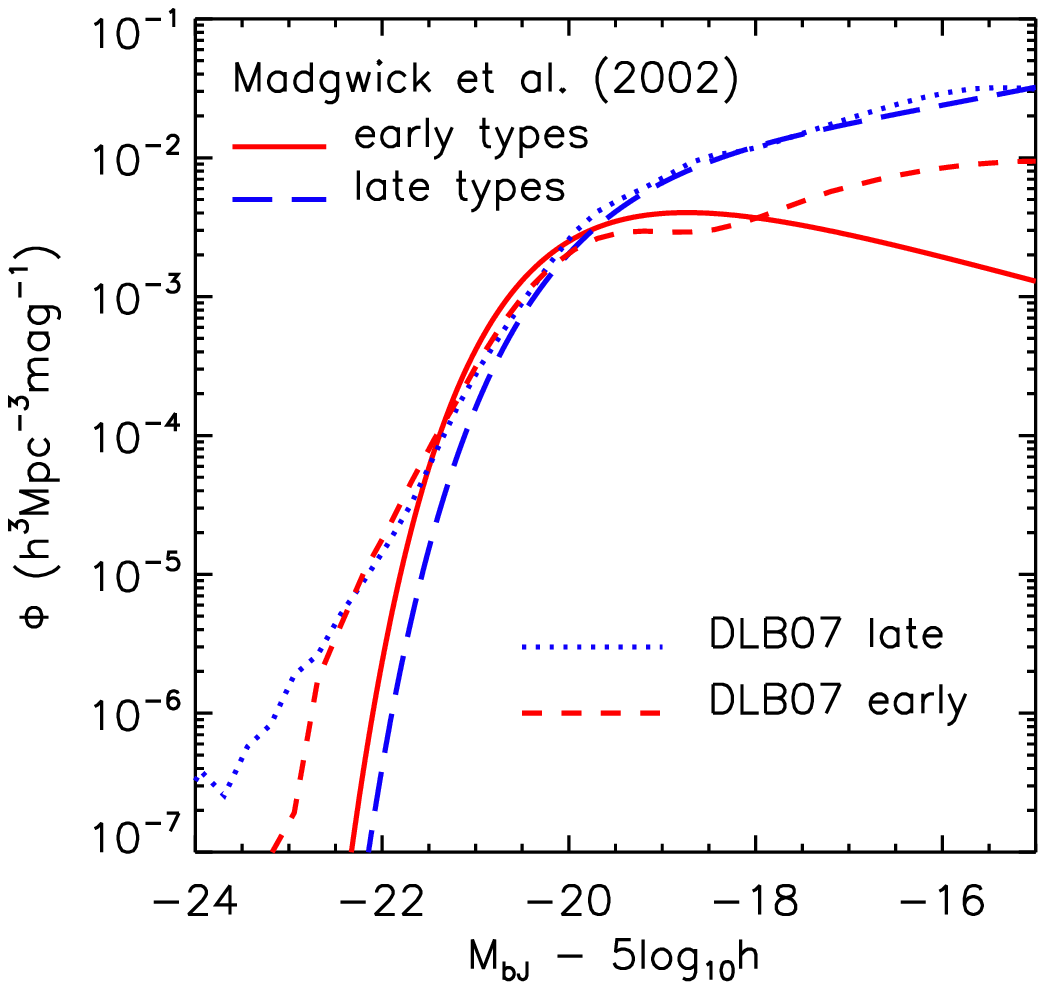}
\includegraphics[width=8.4cm]{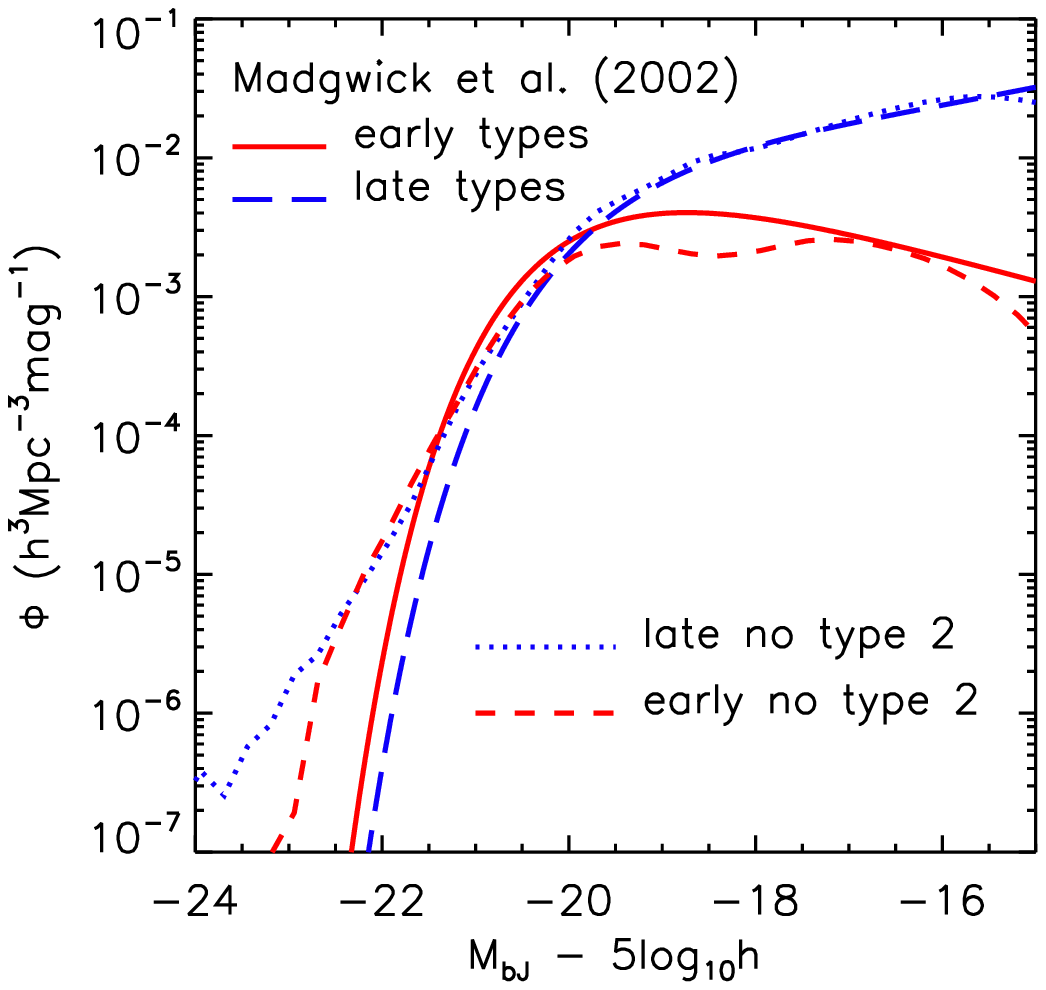}
\caption{The galaxy luminosity function in \bj-band divided by
colour. The upper panel shows results for the \canonical\ model, the
lower panel for the \disruption\ model. In both panels the model
predictions for the red and blue population (red dashed and blue
dotted lines respectively) are compared with 2dFGRS data of
\citet{Madgwick2002} for early- and late-type galaxies (red solid and
blue long-dashed lines respectively).}
\label{fig:2lf}
\end{figure}

The disagreement between the model predictions for the number density
of faint galaxies and observations has been identified in the
past. \citet{Croton2006} showed that this is mainly due to an excess
of early-type, red galaxies. The top panel of Fig.~\ref{fig:2lf}
illustrates this discordance.  The long-dashed and the solid lines in
both panels of Fig.~\ref{fig:2lf} show the luminosity functions of
late- and early-type galaxies respectively for the 2dFGRS data
\citep{Madgwick2002}. The dotted and dashed lines show the predictions
for late- and early-type galaxies respectively for the DLB07
model. The populations of late (blue) and early (red) galaxies are
defined as a function of $B-V$ colour, late galaxies having $B-V <
0.8$ and early galaxies $B-V > 0.8$, respectively.  The lower panel of
Fig.~\ref{fig:2lf} shows the effect of disrupting type 2
galaxies. Whilst the luminosity function of late type galaxies is
hardly affected, the reduction in the number density of faint early
type galaxies is dramatic (more than an order of magnitude) and
brings the model into much closer agreement with the observations.
The luminosity function of early type galaxies shows a dip at $\bj
\sim -19$.  However, this dip can also be seen in the luminosity
function for the canonical model and is not caused by galaxy
disruption.

\subsection{Contribution to the ICL}
\label{sec:massfrac}

\begin{figure}
\centering
\includegraphics[width=8.4cm]{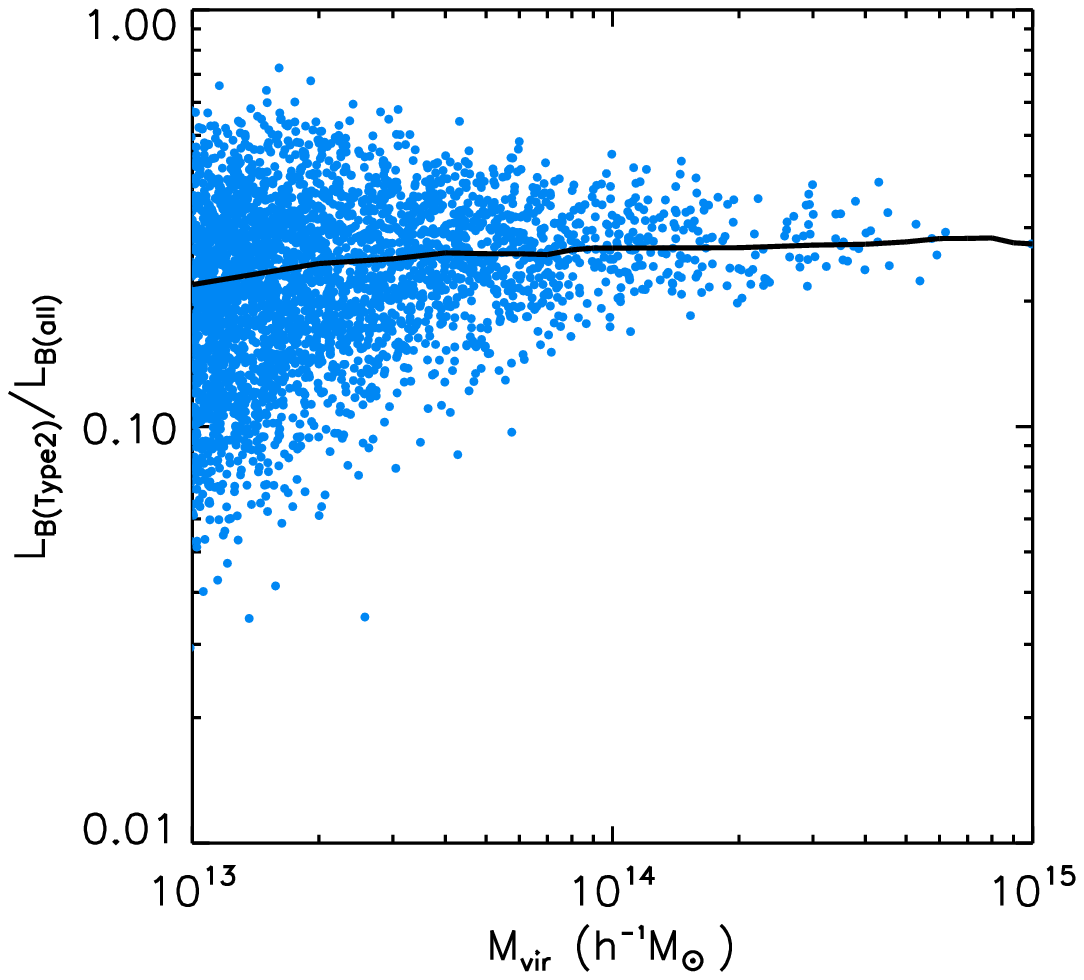}
\caption{Fraction of the luminosity of type 2 galaxies over the total
galaxy luminosity in the $B$-band as a function of virial mass for
the \disruption\ model. The solid line represents the median of
the $L_{\rm B(Type2)}$/ $L_{\rm B(all)}$ distribution.}
\label{fig:massfrac}
\end{figure}

Despite of the uncertainty related to the difficulty in detecting very
faint sources, observations suggest that diffuse intracluster stars
contribute a substantial fraction of the optical light emitted by
clusters and rich groups.  In massive clusters, between 10 and 50 per
cent of the optical luminosity is estimated to be diffuse ICL.  In
Fig.~\ref{fig:massfrac} we show our results for the fraction of
$B-$band luminosity contributed by type 2 galaxies, compared to the
total luminosity of a cluster, which is the sum of the $B$
luminosities of all the galaxies in the cluster, including type 2
galaxies. Results are presented for groups and clusters with $M\vir >
10^{13}\hm \Msun$.  Although there is considerable scatter in the
distribution, our estimated fractions fall well within the observed
range (\citealt{Bernstein1995}; \citealt{Gonzalez2000};
\citealt{Feldmeier2002}; \citealt{Feldmeier2004b};
\citealt{Gonzalez2005}; \citealt{Zibetti2005}), and agree with
the predictions of other semi-analytic models where galaxy disruption
is implemented self-consistently \citep{Purcell2007}.  We find that the
median $B$ luminosity of type 2 galaxies accounts for about 20 per
cent of the total luminosity for haloes with $M\vir \sim 10^{13}\hm
\Msun$, increasing to about $30$ per cent in haloes with $M\vir=
10^{15}\hm \Msun$. This is strong evidence that disruption of dwarf
galaxies is not only helpful in reducing the number density of faint
red galaxies but is actually required by the observations, as there is
no other plausible mechanism for generating the ICL.

\section{Conclusions}
\label{sec:discuss}

In this paper, we have considered the possibility that dwarf galaxies
may become disrupted when subjected to gravitational tidal forces and
repeated galaxy encounters within clusters and rich galaxy groups.
For this purpose, we have introduced an \emph{ad hoc} prescription for
the disruption of galaxies in our analysis of galaxy catalogues
produced by semi-analytic models of galaxy formation. Our simple
prescription considers the stability of substructure within dark
matter haloes.  The merger trees produced by the algorithm {\small
SUBFIND} \citep{Springel2001} can identify dark matter sub-haloes
within a larger structure until they are completely disrupted. We make
the simple assumption that galaxies whose haloes have been disrupted
will themselves disperse, while their stars become part of the
population of intracluster stars responsible for the observed
intracluster light.

Despite the simplicity of this assumption, our results show a
substantial improvement over previous models and indicate that the
inclusion of galaxy disruption is indeed a necessary ingredient of
galaxy formation models.  Our main results can be summarised as
follows:
\begin{itemize}
\item it suppresses the number of red dwarf galaxies, improving the
model predictions for the faint end of the galaxy luminosity function
in \bj-band;
\item correspondingly, it suppresses the number density of the faint,
reddest galaxies in colour-magnitude diagrams;
\item it predicts values for the fraction and colour of intracluster
light in good agreement with observational estimates for clusters and
rich galaxy groups.
\end{itemize}

Our prescription to model the ICL, implemented \emph{a posteriori} in
the publicly available galaxy catalogue of DLB07, assumes that all
type 2 galaxies that have survived merging until $z=0$ are disrupted
and contribute their stars to the population of intracluster stars. By
doing this, the evolution of type 2 galaxies at $z>0$ is not affected.

Although the simplicity of our model is to be valued, the assumption
that halo disruption leads to galaxy disruption is very simplistic.
Hydrodynamical simulations (\citealt{Moore1996};
\citealt{Napolitano2003}; \citealt{Murante2004};
\citealt{Willman2004}; \citealt{Sommer-Larsen2005};
\citealt{Monaco2006}; \citealt{Murante2007}) suggest that tidal
stripping of stars from the outer regions of a galaxy is a gradual
process that takes several passages of the galaxy through the cluster
core and that can be enhanced by previous galaxy encounters in
groups. Furthermore, a galaxy infalling into a cluster might not be
completely disrupted, but might conserve its densest core intact, as
the existence of UCD galaxies seem to suggest.

A natural extension of the results presented here would be to
introduce the halo disruption self-consistently within the
semi-analytic model, as \citet{Monaco2006} have done. This would have
several effects, the main ones being to limit the growth of central
galaxies and to affect the abundance of dwarf galaxies. We intend to
follow this up in future work.

\section*{Acknowledgements}
We would like to thank Ivan Baldry for supplying us with the
observational data for Fig.~\ref{fig:cm}. BMH would also like to thank
Jon Loveday for the support and supervision through the build--up of
this paper. Many thanks to Gerard Lemson for setting up the GAVO
Millennium database and for his continuous efforts on its development
and outreach.  BMH thanks the Portuguese Science and Technology
Foundation (FCT) for financial support through an International Ph.D.
fellowship (grant number SFRH/BD/21497/2005).  SB and PAT are
supported by STFC.  The Millennium simulation was carried out by the
Virgo Consortium at the Max Planck Society in Garching.  Data on the
galaxy population produced by this model, as well as on the parent
halo population, are publicly available at
{\tt http://www.mpa-garching.mpg.de/millennium/}.

\bibliographystyle{mn2e}
\bibliography{paper}

\label{lastpage}

\end{document}